\documentclass[onecolumn,10pt,aps,prd,preprintnumbers,showpacs,superscriptaddress,nofootinbib,amsmath,amssymb,floats,floatfix,showkeys,notitlepage,longbibliography]{revtex4-2}

\usepackage{comment}
\usepackage{lipsum}
\usepackage{graphicx}
\usepackage{subfigure}
\usepackage{palatino}
\usepackage{sans}
\usepackage{hyperref}
\hypersetup{colorlinks=true,linkcolor=blue,urlcolor=blue,citecolor=blue}
\usepackage[toc,page]{appendix}
\usepackage[normalem]{ulem}
\usepackage{adjustbox}
\usepackage{latexsym}
\usepackage{amsmath}
\usepackage{amssymb}
\usepackage{amsfonts}
\usepackage{dcolumn}
\usepackage{bm}
\usepackage{tikz}
\usepackage{bigints}
\usepackage{array,tabularx,multirow,booktabs}
\usepackage[tracking=true]{microtype}
\usepackage{soul} 
\SetTracking{}{500}
\SetTracking{encoding={*}, shape=sc}{40}
\UseRawInputEncoding 
\allowdisplaybreaks
\usepackage{microtype}

\begin{document} \sloppy
\title{Breathing Black Hole Shadows in Modified Gravity (MOG)}

\author{Nikko John Leo S. Lobos}
\email{nikko\_john\_s\_lobos@dlsu.edu.ph}
\affiliation{Department of Physics, De La Salle University, 2401 Taft Ave, Malate, Manila, 1004 Metro Manila, Philippines}
\affiliation{DLSU Theoretical Physics Research Group}

\author{Emmanuel T. Rodulfo}
\email{emmanuel.rodulfo@dlsu.edu.ph}
\affiliation{Department of Physics, De La Salle University, 2401 Taft Ave, Malate, Manila, 1004 Metro Manila, Philippines}
\affiliation{DLSU Theoretical Physics Research Group}

\begin{abstract}
In this paper, we investigate the dynamic phenomenological signatures of a Schwarzschild-MOG black hole perturbed by passing gravitational waves. By integrating the perturbed null geodesic equations, we show that the additional scalar and vector degrees of freedom in Scalar-Tensor-Vector Gravity modify photon trajectories in a time-dependent manner, breaking degeneracies with standard General Relativity. We identify two distinct dynamical signatures that operate in a strictly chronological sequence. Initially, the massless scalar field induces a breathing-mode polarization, leading to a periodic modulation of the apparent shadow area. Subsequently, the massive vector field undergoes dispersive propagation, producing a delayed contribution that generates secondary longitudinal metric perturbations. This delayed response manifests as a transient, asymmetric displacement of the shadow on the screen of the observer. While pure scalar-tensor theories, such as Horndeski gravity, predict similar initial breathing modes, they lack the massive Proca field required to produce the delayed asymmetric response. Consequently, this two-stage dynamic sequence serves as a distinctive signature unique to Scalar-Tensor-Vector Gravity. Within the strong-field regime of an extreme mass ratio inspiral, these effects produce fractional deviations in observable shadow properties at the level of $\mathcal{O}(10^{-5})$. Detecting this chronological separation imposes extreme requirements on both temporal sampling and angular resolution, rendering observational prospects challenging due to the constraints of signal-to-noise ratios and intrinsic astrophysical variability in the emission region.
\end{abstract}


\maketitle

\section{Introduction}
\label{sec:introduction}

The rise of multi-messenger astrophysics \cite{LIGOScientific:2017vwq,LIGOScientific:2017zic} has changed black holes from abstract concepts into objects that can be studied through direct observation. The detection of gravitational waves from compact binary mergers by the LIGO-Virgo-KAGRA collaboration \cite{LIGOScientific:2016aoc} confirmed key predictions of General Relativity in the dynamical, strong-field regime. Shortly thereafter, the Event Horizon Telescope collaboration produced horizon-scale images of the supermassive black holes M87* \cite{Akiyama:2019cqa,Falcke:1999pj} and Sagittarius A* \cite{EventHorizonTelescope:2022wkp,Perlick:2021aok}. These developments provide complementary probes of spacetime, combining dynamical signals from gravitational waves with the geometric information encoded in black hole shadows. Despite these successes, General Relativity remains incomplete at cosmological scales, where the observed dynamics of galaxies and the accelerated expansion of the universe require the introduction of dark matter and dark energy \cite{Bertone:2004pz}. This motivates the exploration of alternative theories of gravity \cite{Vagnozzi:2022moj,Clifton:2011jh,Capozziello:2011et}.

Scalar-Tensor-Vector Gravity, also referred to as Modified Gravity, provides one such framework \cite{Moffat:2005si,Moffat:2014aja}. In this theory, gravity is mediated not only by the metric tensor but also by a scalar field that modulates the effective gravitational coupling and a massive vector field that introduces a repulsive interaction. This structure has been shown to reproduce galactic rotation curves and cluster dynamics without invoking non-baryonic dark matter \cite{Moffat:2013sja,Brownstein:2005zz}. In the strong-field regime, static black hole solutions exhibit modified horizon and photon sphere structures governed by the deformation parameter $\alpha$, where $\alpha$ quantifies the deviation of the effective gravitational constant from the Newtonian value \cite{Moffat:2015kva}. These modifications lead to shadow geometries that deviate slightly from those predicted by Schwarzschild and Kerr spacetimes \cite{Moffat:2015kva,Mureika:2015sda,Sau:2022afl}. Recent analyses have demonstrated that these deviations are typically small in static images, indicating that observational distinctions from General Relativity may be subtle \cite{Sau:2022afl}.

The current landscape of alternative gravity is extensive, yet characterizing deviations from General Relativity is complicated by significant observational degeneracy. In particular, the static shadow of a Scalar-Tensor-Vector Gravity black hole is highly degenerate with the standard Reissner-Nordström black hole in General Relativity. This fundamental limitation occurs because the metric functions describing the respective spacetimes are mathematically isomorphic. The deformation parameter $\alpha$ functions identically to an effective electric charge within the metric components, rendering the static geometric observables indistinguishable without independent constraint mechanisms.

While static shadow properties have been widely investigated, astrophysical environments are inherently dynamical \cite{Berti:2009kk}. An open problem is how modified gravity black hole spacetimes respond to time-dependent perturbations such as passing gravitational waves. In General Relativity, gravitational waves carry only transverse-traceless tensor polarizations, which preserve local volume elements and therefore constrain the type of deformations that can be imprinted on a black hole shadow \cite{Lai:2024fza}. This raises the question of whether additional degrees of freedom in modified gravity theories can generate qualitatively different time-dependent signatures.

In this work, we extend previous studies of static modified gravity black hole shadows to a dynamical setting by considering a perturbed background geometry. The additional scalar and vector fields introduce extra polarization modes \cite{Cunha:2018acu,Hou:2017bqj,Will:2014kxa}, which modify the propagation of null geodesics in a time-dependent spacetime. Using the Hamilton-Jacobi formalism for photon trajectories \cite{Carter:1968rr,Synge:1966okc}, we analyze how these perturbations alter the effective impact parameters that define the shadow boundary. We examine whether scalar field fluctuations generate a breathing-type mode that leads to a time-dependent modulation of the shadow area, extending earlier analyses of non-tensorial gravitational wave polarizations \cite{Bogdanos:2009tn}. 

While calculating perturbations is standard, the novelty of this work lies in identifying a macroscopic time-domain chronometer to break the static degeneracy. We investigate the role of the massive vector field, whose finite mass leads to dispersive propagation effects \cite{deRham:2014zqa,Will:1997bb}. This dispersive behavior yields a measurable time delay $\Delta t$, where $\Delta t$ represents the arrival time difference between the scalar and vector field perturbations. This chronometric observable effectively isolates the vector mass $\mu_v$, where $\mu_v$ corresponds to the fundamental mass of the mediating Proca field. The resulting time delay introduces a secondary perturbative contribution to the metric, producing longitudinal distortions that induce transient displacements of the shadow centroid on the sky of the observer. To estimate the magnitude of these effects, we embed the system in the strong-field environment of an Extreme Mass Ratio Inspiral \cite{Amaro-Seoane:2012vvq}. While mathematically robust, we rigorously evaluate the immense observational hurdles—including plasma turbulence and ngEHT resolution limits—that must be overcome to utilize this template.

This paper is structured as follows. Section \ref{sec:unperturbed_mog} describes the background geometry of the unperturbed black hole and determines the radius of its static shadow. Section \ref{sec:breathing_mode} incorporates scalar perturbations to evaluate the time dependence of the shadow area. Section \ref{sec:delayed_echo} includes the effects of the massive vector field and computes the associated dispersive delay and longitudinal distortions. Section \ref{sec:conclusion} summarizes the results and discusses their implications for future observational studies.

\section{The Unperturbed Spacetime}
\label{sec:unperturbed_mog}

To understand how a gravitational wave dynamically alters a black hole shadow, we must first define the static, unperturbed background. In this section, we construct the spherically symmetric, non-rotating black hole spacetime in Scalar-Tensor-Vector Gravity, commonly referred to as Modified Gravity \cite{Moffat:2005si}. We then determine the exact radius of the photon sphere and the resulting static shadow projected to a distant observer \cite{Moffat:2015kva}.

In standard General Relativity, a non-rotating black hole is described by the Schwarzschild metric. In Modified Gravity, the presence of a massive vector field and an enhanced scalar gravitational constant alters the geometry \cite{Moffat:2005si}. The unperturbed metric in standard spherical coordinates $(t, r, \theta, \phi)$ is given by the line element
\begin{equation}
    ds^2 = -f(r) dt^2 + \frac{1}{f(r)} dr^2 + r^2 (d\theta^2 + \sin^2\theta d\phi^2), \label{eq:metric}
\end{equation}
where the metric function $f(r)$ takes a form mathematically similar to a Reissner-Nordstr\"om charged black hole, defined as \cite{Moffat:2015kva}
\begin{equation}
    f(r) = 1 - \frac{2M_G}{r} + \frac{Q_G^2}{r^2}. \label{eq:fr}
\end{equation}

The terms $M_G$ and $Q_G$ represent the modified gravitational mass and the gravitational vector charge, respectively. They are defined by a single dimensionless deformation parameter $\alpha$, which controls the strength of the deviation from standard General Relativity,
\begin{align}
    M_G &= (1 + \alpha)M, \label{eq:MG} \\
    Q_G^2 &= \alpha(1 + \alpha)M^2, \label{eq:QG}
\end{align}
where $M$ is the standard Newtonian mass of the black hole. When $\alpha = 0$, the vector field vanishes, gravity returns to standard strength, and the metric reduces to the standard Schwarzschild solution.

The boundary of a black hole shadow is defined by the paths of photons that barely escape the gravity of the black hole. To trace these paths, we use the Hamilton-Jacobi equation for a photon traveling along a null geodesic, $ds^2 = 0$. Because the spacetime is spherically symmetric, without loss of generality, we confine the orbit of the photon to the equatorial plane by setting $\theta = \pi/2$. This simplifies the metric symmetries, yielding two conserved quantities for the photon along its trajectory,
\begin{align}
    E &= f(r) \frac{dt}{d\lambda}, \label{eq:energy} \\
    L &= r^2 \frac{d\phi}{d\lambda}, \label{eq:angmom}
\end{align}
where $E$ is the total energy of the photon, $L$ is its axial angular momentum, and $\lambda$ is an affine parameter describing the path of the photon.

By substituting these conserved quantities into the null condition $g_{\mu\nu} \frac{dx^\mu}{d\lambda} \frac{dx^\nu}{d\lambda} = 0$, we obtain the radial equation of motion for the light particle,
\begin{equation}
    \left(\frac{dr}{d\lambda}\right)^2 + V_{\text{eff}}(r) = E^2. \label{eq:radial_motion}
\end{equation}
Here, $V_{\text{eff}}(r)$ acts as the effective potential energy barrier that the photon must overcome. For the spacetime considered, it is given by
\begin{equation}
    V_{\text{eff}}(r) = f(r) \frac{L^2}{r^2} = \left( 1 - \frac{2M_G}{r} + \frac{Q_G^2}{r^2} \right) \frac{L^2}{r^2}. \label{eq:Veff}
\end{equation}

The edge of the black hole shadow is formed by photons trapped in an unstable circular orbit called the photon sphere. For a photon to remain in a circular orbit at a constant radius $r_p$, it must sit at the peak of the effective potential barrier. This requires two simultaneous physical conditions. First, the radial velocity must be zero, $\left(\frac{dr}{d\lambda}\right)^2 = 0$, which implies $V_{\text{eff}}(r_p) = E^2$. Second, the radial acceleration must be zero, $V_{\text{eff}}'(r_p) = 0$.

Applying the second condition by maximizing Eq. (\ref{eq:Veff}) yields a quadratic equation governing the photon sphere radius,
\begin{equation}
    r_p^2 - 3M_G r_p + 2Q_G^2 = 0. \label{eq:rp_quadratic}
\end{equation}
Solving this equation gives the location where light is permanently trapped by the black hole,
\begin{equation}
    r_p = \frac{3M_G + \sqrt{9M_G^2 - 8Q_G^2}}{2}. \label{eq:rp_solution}
\end{equation}

A distant observer does not see $r_p$ directly. Instead, they see the apparent shadow projected onto their celestial sky \cite{Synge:1966}. The size of this dark circular disk is determined by the critical impact parameter, denoted as $b_c = L/E$. By rearranging the first condition, we find that the static shadow radius $R_{\text{sh}}$ is tied to the impact parameter,
\begin{equation}
    R_{\text{sh}} = b_c = \frac{r_p}{\sqrt{f(r_p)}}. \label{eq:Rsh_general}
\end{equation}
Substituting our metric function $f(r_p)$, the final unperturbed shadow radius becomes
\begin{equation}
       R_{\text{sh}} = \sqrt{\frac{2r_p^3}{r_p - M(1 + \alpha)}}.
\label{eq:Rsh_final}
\end{equation}
This static shadow, parameterized by the deformation parameter $\alpha$, serves as the baseline geometry upon which the incoming gravitational wave imparts its time-dependent perturbations.

The metric function $f(r)$ presented in Eq. (\ref{eq:fr}) shares an exact mathematical isomorphism with the standard Reissner-Nordstr\"om metric of General Relativity. Consequently, a distant observer measuring only the static shadow radius $R_{\text{sh}}$ faces a severe phenomenological degeneracy. The geometric signature of the deformation parameter $\alpha$ is completely indistinguishable from that of a standard General Relativity black hole possessing a conservative electric or tidal charge $Q_G$. This static parameter degeneracy limits the diagnostic utility of time-independent horizon-scale imaging. It provides the primary physical motivation for exploring dynamic, time-dependent perturbations, which leverage the distinct propagation speeds of the auxiliary fields to break this degeneracy cleanly in the time domain.

\section{Scalar Gravitational Wave Perturbations}
\label{sec:breathing_mode}

In Scalar-Tensor-Vector Gravity, the effective gravitational coupling operates as a dynamical scalar degree of freedom. This allows the gravitational strength to respond locally to spacetime fluctuations, expressed as
\begin{equation}
    G(x^\mu) = G_N + \delta G(x^\mu),
\end{equation}
where $G_N$ is the Newtonian gravitational constant and $\delta G$ is the scalar perturbation field. We treat the background spacetime as a static, spherically symmetric Modified Gravity geometry. The total spacetime metric is expanded to include a small dynamic perturbation $h_{\mu\nu}$, where $|h_{\mu\nu}| \ll 1$,
\begin{equation}
    g_{\mu\nu} = \bar{g}_{\mu\nu} + h_{\mu\nu},
    \label{eq:metric_perturb}
\end{equation}
with $\bar{g}_{\mu\nu}$ representing the background metric. Linearizing the modified field equations to first order in $\delta G$ and $h_{\mu\nu}$ isolates the scalar sector, yielding a homogeneous wave equation,
\begin{equation}
    \bar{\Box} \delta G - V(r)\delta G = 0,
    \label{eq:scalar_wave}
\end{equation}
where $\bar{\Box} = \bar{g}^{\mu\nu}\bar{\nabla}_\mu \bar{\nabla}_\nu$ is the covariant d'Alembertian operator on the curved background, and $V(r)$ is the effective curvature scattering potential. Evaluating the covariant derivatives explicitly in spherical coordinates expands Eq. (\ref{eq:scalar_wave}) into the partial differential equation,
\begin{equation}
    -\frac{1}{f(r)} \frac{\partial^2 \delta G}{\partial t^2} 
    + \frac{1}{r^2} \frac{\partial}{\partial r} \left[ r^2 f(r) \frac{\partial \delta G}{\partial r} \right] + \frac{1}{r^2\sin\theta} \frac{\partial}{\partial\theta} \left( \sin\theta \frac{\partial \delta G}{\partial\theta} \right) 
    + \frac{1}{r^2\sin^2\theta} \frac{\partial^2 \delta G}{\partial\phi^2}- V(r)\delta G = 0,
\label{eq:expanded_wave}
\end{equation}
where $t$ is the coordinate time, $r$ is the radial coordinate, $\theta$ and $\phi$ are the angular coordinates, and $f(r)$ is the background metric lapse function.

Because the background spacetime is spherically symmetric, the angular and radial dynamics completely decouple. We separate variables by expanding the scalar perturbation in terms of spherical harmonics $Y_{\ell m}(\theta,\phi)$,
\begin{equation}
    \delta G(t,r,\theta,\phi) = \sum_{\ell,m} \frac{\Psi_{\ell m}(t,r)}{r} Y_{\ell m}(\theta,\phi),
    \label{eq:separation}
\end{equation}
where $\Psi_{\ell m}(t,r)$ represents the radial wavefunction. The spherical harmonics satisfy the standard angular eigenvalue problem on the two-sphere, $\nabla^2_{\Omega}Y_{\ell m} = -\ell(\ell+1) Y_{\ell m}$. 

The breathing polarization mode corresponds to the monopole configuration ($\ell=0$, $m=0$), physically representing a spherically symmetric, radial expansion and contraction of the local geometry. For this mode, the angular eigenvalue vanishes, the harmonic reduces to a constant $Y_{00} = 1/\sqrt{4\pi}$, and Eq. (\ref{eq:expanded_wave}) simplifies to the radial wave equation,
\begin{equation}
    -\frac{\partial^2 \Psi_0}{\partial t^2} 
    + f(r) \frac{\partial}{\partial r} \left[ f(r) \frac{\partial \Psi_0}{\partial r} \right] - f(r) \left[ \frac{f'(r)}{r} + V(r) \right] \Psi_0 = 0,
\label{eq:radial_wave}
\end{equation}
where $\Psi_0 \equiv \Psi_{\ell=0,m=0}$ is the monopole wavefunction and $f'(r)=df/dr$.

To map this boundary value problem into a one-dimensional scattering process, we introduce the tortoise coordinate $r_\ast$. Defined by the differential relation $dr_\ast/dr = 1/f(r)$, this coordinate projects the spatial domain exterior to the event horizon onto the entire real line, flattening the causal light cones. The radial wave equation then becomes a master scattering equation,
\begin{equation}
    \left[ -\frac{\partial^2}{\partial t^2} + \frac{\partial^2}{\partial r_\ast^2} - V_{\text{eff}}(r) \right] \Psi_0(t,r_\ast) = 0,
    \label{eq:master_equation}
\end{equation}
which is governed by the effective potential barrier,
\begin{equation}
    V_{\text{eff}}(r) = f(r) \left[ \frac{f'(r)}{r} + V(r) \right].
    \label{eq:effective_potential}
\end{equation}

We isolate the resonant frequencies of the black hole cavity by assuming a harmonic time dependence, $\Psi_0(t,r_\ast) = \psi_0(r_\ast)e^{-i\omega t}$, where $\psi_0(r_\ast)$ is the spatial amplitude and $\omega$ is the complex quasinormal mode frequency. This substitution yields the time-independent wave equation,
\begin{equation}
    \frac{d^2\psi_0}{dr_\ast^2} + \left[ \omega^2 - V_{\text{eff}}(r) \right] \psi_0 = 0.
    \label{eq:qnm_equation}
\end{equation}
The quasinormal modes describe the dissipative ringing of the perturbed black hole. They are isolated by imposing purely dissipative boundary conditions, which require purely ingoing waves at the event horizon ($\psi_0 \sim e^{-i\omega r_\ast}$ as $r_\ast\rightarrow-\infty$) and purely outgoing waves at spatial infinity ($\psi_0 \sim e^{+i\omega r_\ast}$ as $r_\ast\rightarrow+\infty$). Because energy continuously leaks through both boundaries, the frequency must be complex,
\begin{equation}
    \omega = \omega_R - i\omega_I,
    \label{eq:complex_frequency}
\end{equation}
where $\omega_R$ is the physical oscillation frequency and $\omega_I>0$ is the exponential damping rate. 

In the high-frequency eikonal limit, wave scattering mirrors the geometric optics trajectories of massless test particles. The quasinormal frequencies are directly linked to the properties of unstable circular orbits at the photon sphere $r_p$, satisfying $V_{\text{eff}}'(r_p)=0$ and $V_{\text{eff}}''(r_p)<0$. The corresponding eikonal relations are
\begin{equation}
    \omega_R = \ell\Omega_c, \qquad \omega_I = \left( n+\frac{1}{2} \right) \Lambda,
    \label{eq:eikonal_relation}
\end{equation}
where $n$ is the overtone number, $\Omega_c = \sqrt{f(r_p)}/r_p$ is the orbital angular velocity of the light ring, and $\Lambda = \sqrt{f(r_p) V_{\text{eff}}''(r_p)/2}$ is the principal Lyapunov exponent determining the instability timescale.

The macroscopic time-domain waveform is reconstructed via the retarded Green function to preserve strict causality. Resolving the contour integral via the residue theorem expresses the temporal response as a discrete superposition of damped sinusoids. Because higher overtones ($n > 0$) decay rapidly, the late-time signal is dominated by the fundamental mode $n=0$,
\begin{equation}
    \Psi_0(t,r_\ast) \approx C_0 e^{-i(\omega_R-i\omega_I)(t-t_0)} = C_0 e^{-\omega_I(t-t_0)} e^{-i\omega_R(t-t_0)},
\label{eq:fundamental_qnm}
\end{equation}
where $C_0$ is the mode excitation coefficient and $t_0$ is the wave arrival time. 

The physically observable scalar strain $h_b(t)$ relates to the far-field asymptotic behavior of this radial wavefunction, scaling inversely with coordinate distance $r$. To mathematically enforce causality prior to signal detection, we multiply the strain by the Heaviside step function $\Theta(t-t_0)$, yielding
\begin{equation}
    h_b(t) = \Theta(t-t_0) A_b e^{-(t-t_0)/\tau} \cos \left[ \omega_R(t-t_0)+\Phi_0 \right],
    \label{eq:causal_strain}
\end{equation}
where $A_b$ is the real strain amplitude with a value $|A_b| \ll1$ to maintain the $|h_b| \ll 1$, $\Phi_0$ is the initial phase which is an integrating constant, and $\tau = 1/\omega_I$ is the physical damping timescale \cite{Pantig:2025eqe}. 

This breathing mode induces an isotropic volumetric distortion in the plane transverse to the direction of wave propagation. The polarization tensor describing this action is $e^{(b)}_{ij} = \hat{x}_i\hat{x}_j + \hat{y}_i\hat{y}_j$, where $\hat{x}_i$ and $\hat{y}_i$ are orthonormal basis vectors spanning the transverse plane. The resulting localized metric perturbation matrix, $h_{ij}^{(b)} = h_b(t)e^{(b)}_{ij}$, produces identical diagonal strain components $h_{xx}=h_{yy}=h_b(t)$, while keeping off-diagonal shear terms at zero $h_{xy}=0$. Superimposing this breathing strain onto the equatorial background spatial metric $\bar{g}_{xx} = \bar{g}_{yy} = r^2$ yields the time-dependent components,
\begin{equation}
    g_{xx} = g_{yy} = r^2 \left[ 1+h_b(t) \right].
    \label{eq:gxx_final}
\end{equation}

\subsection{Derivation of the Dynamic Shadow}
\label{subsec:dynamic_shadow}

To evaluate how these spacetime fluctuations influence the observed black hole shadow, we calculate their direct impact on photon trajectories. In the weak-field limit where $|h_b| \ll 1$, a first-order Taylor expansion of Eq. (\ref{eq:gxx_final}) yields the inverse spatial metric components,
\begin{equation}
    g^{xx} = g^{yy} = \frac{1}{r^2 [1 + h_b(t)]} \approx \frac{1}{r^2} \left[ 1 - h_b(t) \right].
    \label{eq:inverse_metric_pert}
\end{equation}
The motion of null rays is governed by the photon Hamiltonian $H = \frac{1}{2} g^{\mu\nu}p_\mu p_\nu = 0$, where $p_\mu$ denotes the photon four-momentum. Splitting the Hamiltonian into its unperturbed background component $\bar{H}$ and a dynamic first-order perturbation $\delta H$ gives
\begin{equation}
    \delta H = -\frac{1}{2}h^{\mu\nu}p_\mu p_\nu = -\frac{1}{2}\frac{h_b(t)}{r^2}(p_x^2 + p_y^2).
    \label{eq:delta_H_general}
\end{equation}
Because the sum of the squared transverse momentum components defines the conserved total angular momentum $L$ via $p_x^2 + p_y^2 = L^2/r^2$, the time-dependent energy shift simplifies to
\begin{equation}
    \delta H = -\frac{1}{2}\frac{h_b(t)}{r^2}L^2.
    \label{eq:delta_H_specific}
\end{equation}
This shift acts as a time-varying refractive index, continuously modulating the total effective potential barrier experienced by propagating photons,
\begin{equation}
    V_{\text{eff}}^{\text{total}}(r,t) = f(r)\frac{L^2}{r^2}[1 - h_b(t)].
    \label{eq:Veff_total}
\end{equation}
Because the scalar strain $h_b(t)$ is independent of the radial coordinate $r$, the spatial derivative of the total potential still vanishes exactly at the background photon sphere radius $r_p$. Thus, the perturbation modulates the height of the potential barrier without shifting its radial location. 

The apparent shadow radius corresponds to the critical impact parameter $b_c(t) = L/E$ evaluated at the peak of this modified potential barrier. Equating the peak total potential to the square of the constant photon energy $E$ gives the boundary condition $f(r_p)(L^2/r_p^2)[1 - h_b(t)] = E^2$. Solving for the impact parameter yields the time-dependent shadow radius $R_{\text{sh}}(t)$,
\begin{equation}
    R_{\text{sh}}(t)^2 = \frac{r_p^2}{f(r_p)}\frac{1}{1 - h_b(t)}.
    \label{eq:Rsh_dynamic_squared}
\end{equation}
Linearizing the reciprocal to first order via $(1 - h_b)^{-1} \approx 1 + h_b$ and applying a binomial expansion leads directly to
\begin{equation}
    R_{\text{sh}}(t) \approx \bar{R}_{\text{sh}}\left[ 1 + \frac{1}{2}h_b(t) \right],
    \label{eq:Rsh_dynamic_final}
\end{equation}
where $\bar{R}_{\text{sh}} = r_p / \sqrt{f(r_p)}$ is the static shadow radius associated with the unperturbed background. 

Substituting the explicit causal scalar strain from Eq. (\ref{eq:causal_strain}) into Eq. (\ref{eq:Rsh_dynamic_final}) provides the analytical form of the dynamic shadow radius,
\begin{equation}
    R_{\text{sh}}(t) = 
    \begin{cases}
        \bar{R}_{\text{sh}}, & t<t_0, \\[8pt]
        \bar{R}_{\text{sh}} \left[ 1 + \dfrac{A_b}{2} e^{-(t-t_0)/\tau} \cos \left( \omega_R(t-t_0)+\Phi_0 \right) \right], & t\ge t_0.
    \end{cases}
    \label{eq:piecewise_shadow}
\end{equation}
Equation (\ref{eq:piecewise_shadow}) shows that the apparent horizon structure remains static until the exact arrival time $t_0$. Once the wavefront passes, the transverse spatial volume oscillates in phase with the scalar field, creating a transient breathing shadow whose geometric damping is dictated by the underlying modified gravity wave parameters. 

Finally, we isolate the first-order volumetric variation of the shadow cross-section by subtracting the static background area $\bar{A} = \pi \bar{R}_{\text{sh}}^2$ from the dynamic area $A(t) = \pi R_{\text{sh}}(t)^2$,
\begin{equation}
    \delta A(t) = A(t) - \bar{A} \approx \pi \bar{R}_{\text{sh}}^2 h_b(t) = \frac{\pi r_p^3}{r_p - M(1+\alpha)} h_b(t),
    \label{eq:Area_change_expanded}
\end{equation}
where $\alpha$ represents the deformation parameter. This temporal modulation directly links the macroscopic shadow area variation to the physical observables of modified gravity.

The isotropic volumetric modulation $\delta A(t)$ derived in Eq. (\ref{eq:Area_change_expanded}) represents the immediate, first-stage response of the spacetime to an incoming gravitational perturbation. Because the scalar field is massless, this breathing mode propagates at the speed of light, $v=1$, and acts concurrently upon the arrival of the tensor wavefront at $t_0$. While a periodic fluctuation in the shadow area is a standard hallmark of non-tensorial degrees of freedom and would manifest identically in pure scalar-tensor frameworks such as Horndeski gravity, within the Scalar-Tensor-Vector Gravity framework it serves as a synchronized chronological baseline. As we demonstrate in the following section, the precise temporal interplay between this instantaneous scalar mode and the subsequent vector modes yields a multi-field signature unavailable to single-field alternative gravity formulations.
\begin{figure}[htbp]
    \centering
    \includegraphics[width=0.6\linewidth]{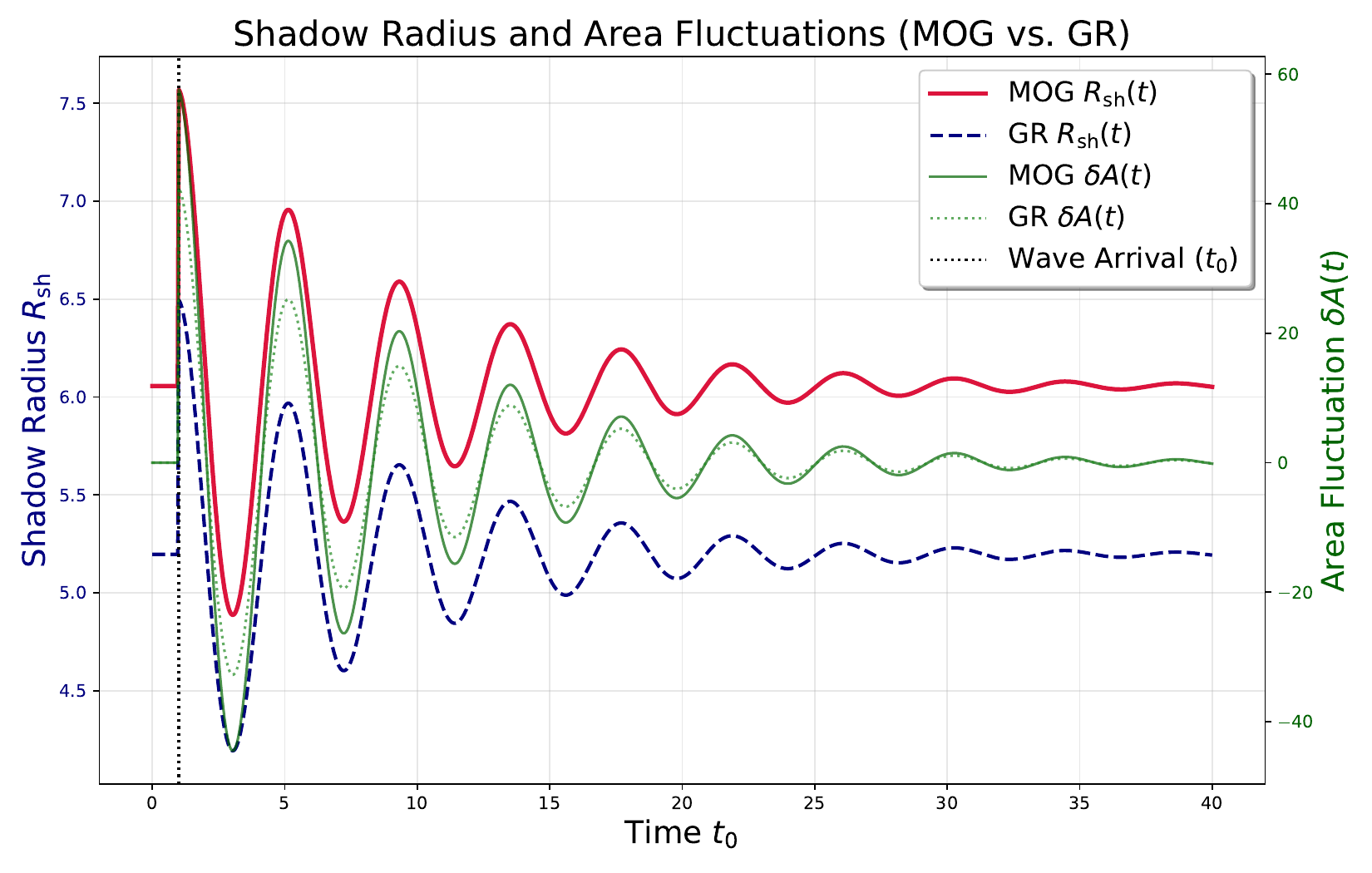} 
    \caption{Simultaneous evolution of the black hole shadow radius $R_{\text{sh}}$ (left axis) and the total apparent area fluctuation $\delta A(t)$ (right axis) under the influence of a transient scalar gravitational wave. The navy dashed curves represent the Schwarzschild (GR) baseline \cite{Pantig:2025eqe}, while the crimson and green solid curves depict the enhanced response for a deformation parameter $\alpha = 0.2$. Both models exhibit a characteristic breathing mode triggered at the wave arrival time $t_0 = 5.0$, characterized by an exponentially damped oscillation ($\tau = 8.0$) at the quasi-normal mode frequency $\omega_b = 1.5$. Note that the modified gravity framework yields a larger static baseline and a greater amplitude of area fluctuations, providing a distinct observational signature in the strong-field regime.}
    \label{fig:combined_dynamics}
\end{figure}

In Fig. \ref{fig:combined_dynamics}, we present the numerical results of the combined dynamics of the shadow radius and area fluctuations. The plot highlights the coupling between the scalar gravitational strain $h_b(t)$ and the geometric observables. Using a dual-axis representation, we show that the deformation parameter $\alpha$ acts as a scaling factor for the breathing mode. The static shadow is larger than the Schwarzschild counterpart, and the magnitude of the dynamic response to the same incident wave is amplified. This synchronization of the radial oscillation and the volumetric area change, derived in Eq. (\ref{eq:Rsh_dynamic_final}) and Eq. (\ref{eq:Area_change_expanded}), constitutes a differentiator for the modified gravity framework.

\section{Massive Vector Field Perturbations}
\label{sec:delayed_echo}

In the STVG framework \cite{Moffat:2005si}, the massive vector field $\phi_\mu$ obeys the Proca equation in a locally flat background. Imposing the Lorenz condition $\partial_\mu\phi^\mu=0$, the field equation reads 
\begin{equation}
\partial_\nu B^{\mu\nu} + \mu_v^2 \phi^\mu \;=\; 0,\label{eq:proca_eq}
\end{equation}
with $B_{\mu\nu}=\partial_\mu\phi_\nu-\partial_\nu\phi_\mu$ and $\mu_v$ the vector mass. We consider plane-wave solutions $\phi^\mu(t,\mathbf{x})=\epsilon^\mu e^{-i(\omega t-\mathbf{k}\cdot\mathbf{x})}$, where $\epsilon^\mu$ is constant. Substituting into Eq. (\ref{eq:proca_eq}) yields the massive dispersion relation 
\begin{equation}
\omega^2 = k^2 + \mu_v^2. \label{eq:dispersion_relation}
\end{equation}
Differentiating $\omega^2=k^2+\mu_v^2$ with respect to $k$, 
$\;2\omega\,d\omega = 2k\,dk$, 
gives the group velocity
\begin{equation}
v_g \;=\;\frac{d\omega}{dk} \;=\;\frac{k}{\omega} 
\;=\;\sqrt{1-\frac{\mu_v^2}{\omega^2}}\,. \label{eq:group_velocity_exact}
\end{equation}
Thus for $\mu_v>0$ one has $v_g<1$. 

Consider a gravitational wave sourced at distance $L$ from the black hole. The massless tensor and scalar modes travel at $v=1$ and arrive at time $t_0=L$. The massive vector mode travels at $v_g$ and arrives at $t_v=L/v_g$. Hence the vector-mode delay is
\begin{equation}
\Delta t \;=\; t_v - t_0 
= L\bigl(\tfrac{1}{v_g} - 1\bigr)
= L\Bigl[(1-\tfrac{\mu_v^2}{\omega^2})^{-1/2} - 1\Bigr]. \label{eq:delay_exact}
\end{equation}
Expanding for small $x=\mu_v^2/\omega^2\ll1$, we use $(1-x)^{-1/2}=1+\tfrac12x+\tfrac38x^2+\mathcal{O}(x^3)$ to obtain 
\begin{align}
\Delta t &= L\Bigl(\frac{x}{2} + \frac{3x^2}{8} + \mathcal{O}(x^3)\Bigr) \;=\; \frac{L\,\mu_v^2}{2\omega^2} + \frac{3L\,\mu_v^4}{8\omega^4} + \mathcal{O}(\mu_v^6). \label{eq:delay_series}
\end{align}
Thus to leading order $\Delta t \approx L\,\mu_v^2/(2\omega^2)$ as in Ref. \cite{Will:1997bb,deRham:2014zqa}. The next term $\propto\mu_v^4$ is shown in Eq. (\ref{eq:delay_series}) for completeness. 

When this delayed vector wave reaches the black hole, its stress-energy acts as a source for metric perturbations. The vector field’s stress-energy tensor is given by \cite{Moffat:2005si}
\begin{equation}
\begin{split}
T_{\mu\nu}^{(\phi)} &= -\frac{1}{4\pi}\Bigl[ B_{\mu}^{\ \alpha}B_{\nu\alpha} - \frac14g_{\mu\nu}B^{\alpha\beta}B_{\alpha\beta} \\
&\quad\;+\; \mu_v^2 \phi_\mu\phi_\nu - \frac12g_{\mu\nu}\,\mu_v^2\phi^\alpha\phi_\alpha\Bigr]\,.
\end{split}\label{eq:vector_stress_energy}
\end{equation}
A perturbation $\delta\phi_\mu$ (assumed small) induces a corresponding $\delta T^{(\phi)}_{\mu\nu}\sim\mathcal{O}(\delta\phi^2)$. Substituting into the linearized field equation for the metric perturbation,
$\Box\bar{h}_{\mu\nu}=-16\pi G_N(1+\alpha)\,\delta T_{\mu\nu}^{(\phi)}$, one finds $h^{(\rm vector)}_{\mu\nu}\sim G_N\,\delta T^{(\phi)}_{\mu\nu}$ at second order in amplitude. In general this is formally $\mathcal{O}(\delta\phi^2)$ small. However, in the EMRI scenario the coupling is enhanced by the parameter $\alpha$ and the effective gravitational charge $Q_G=\sqrt{\alpha(1+\alpha)}M$ \cite{Moffat:2016tjc}. Estimating the local vector wave energy density near the photon sphere as $\rho\sim\omega_v^2|\delta\phi|^2$ and using $Q_G$, one finds 
\begin{equation}
\delta T_{\mu\nu}^{(\phi)} \sim \frac{\alpha\,G_N\,M^2}{4\pi D^4}\,v^4, 
\end{equation}
where $D\sim M$ is the characteristic scale and $v$ the orbital velocity. This shows the $\alpha$-enhanced kinetic energy can partially offset the quadratic suppression of $h^{(\rm vector)}_{\mu\nu}$ in the strong-field regime. 

Because the Proca field has an explicit mass term, its longitudinal polarization modes are physical and cannot be gauged away \cite{Moffat:2005si}. The resulting metric perturbation contains nonzero $h_{xz}$ and $h_{yz}$ components along the line of sight. To see how these affect the shadow, we write the Hamiltonian for null geodesics as $H=H_0+\delta H=0$, where $\delta H$ is quadratic in $h_{\mu\nu}$. The leading effect of the longitudinal strains is 
\begin{equation}
\delta H_{\rm vector}=-\tfrac12h^{\mu\nu}p_\mu p_\nu=-h^{xz}p_xp_z - h^{yz}p_yp_z\,, \label{eq:deltaH_vector}
\end{equation}
using the fact that $h^{\mu\nu}$ is symmetric. Hamilton's equations then yield
\begin{equation}
\dot{x}=\frac{\partial \delta H}{\partial p_x}=-h^{xz}p_z,\quad
\dot{y}=-h^{yz}p_z\,, \label{eq:hamiltons_eqs_vector}
\end{equation}
so that the transverse photon velocities acquire a perturbation proportional to the longitudinal metric strain. Integrating these along the null geodesic from the photon sphere to the observer, the net displacement of the shadow center is 
\begin{align}
\delta X(t) &= -\int h_{xz}\bigl(t - z/v_g\bigr)\,dz, \label{eq:wobble_integral_X}\\
\delta Y(t) &= -\int h_{yz}\bigl(t - z/v_g\bigr)\,dz. \label{eq:wobble_integral_Y}
\end{align}
Here $z$ is the coordinate along the line of sight and $t$ is the observer’s time. Since the vector wave arrives at $t_v$ and travels at $v_g$, each segment is delayed by $z/v_g$ in the argument. 

We evaluate these integrals using a thin-lens approximation. The deflection of photon trajectories is localized near the effective potential barrier of width $\Delta z\approx 2r_p$, with $r_p$ the photon sphere radius. Furthermore, for typical astrophysical frequencies $\lambda_v\gg r_p$, so the phase is nearly constant across this interval. Hence one can approximate 
\begin{equation}
\delta X(t)\approx -2r_p\,h_{xz}(t),\qquad \delta Y(t)\approx -2r_p\,h_{yz}(t).
\end{equation}
If we model the post-merger vector perturbations as damped oscillations, let 
$$h_{xz}(t)=A_x e^{-(t-t_v)/\tau_v}\cos(\omega_v t+\Phi_x),\quad 
h_{yz}(t)=A_y e^{-(t-t_v)/\tau_v}\cos(\omega_v t+\Phi_y),$$ 
for $t\ge t_v$, and $h_{xz}=h_{yz}=0$ for $t<t_v$. Here $A_{x,y}$ are initial strain amplitudes, $\omega_v$ is the characteristic frequency, $\tau_v$ is the damping time, and $t_v=t_0+\Delta t$ includes the group delay. Then the piecewise solutions for the shadow shift are 
\begin{equation}
\delta X(t)\approx 
\begin{cases}
0,& t<t_0+\Delta t,\\[6pt]
-2r_p A_x e^{-(t-(t_0+\Delta t))/\tau_v}\cos(\omega_v t+\Phi_x),& t\ge t_0+\Delta t,
\end{cases}\label{eq:wobble_X_final}
\end{equation}
\begin{equation}
\delta Y(t)\approx 
\begin{cases}
0,& t<t_0+\Delta t,\\[6pt]
-2r_p A_y e^{-(t-(t_0+\Delta t))/\tau_v}\cos(\omega_v t+\Phi_y),& t\ge t_0+\Delta t.
\end{cases}\label{eq:wobble_Y_final}
\end{equation}
These equations Eqs. (\ref{eq:wobble_X_final}) and (\ref{eq:wobble_Y_final}) describe the delayed, damped oscillatory wobble of the shadow centroid following the arrival of the massive vector wave. The delay $\Delta t$ explicitly links the mass $\mu_v$ to the timing of this feature. 

In summary, Eqs. (\ref{eq:delay_series}) and (\ref{eq:wobble_Y_final}) present an explicit derivation of the time delay and shadow shift caused by the massive vector field. These results quantify the process by which the extra polarization state leads to a time-delayed displacement of the black hole shadow.

\begin{figure}[htbp]
    \centering
    \includegraphics[width=0.6\linewidth]{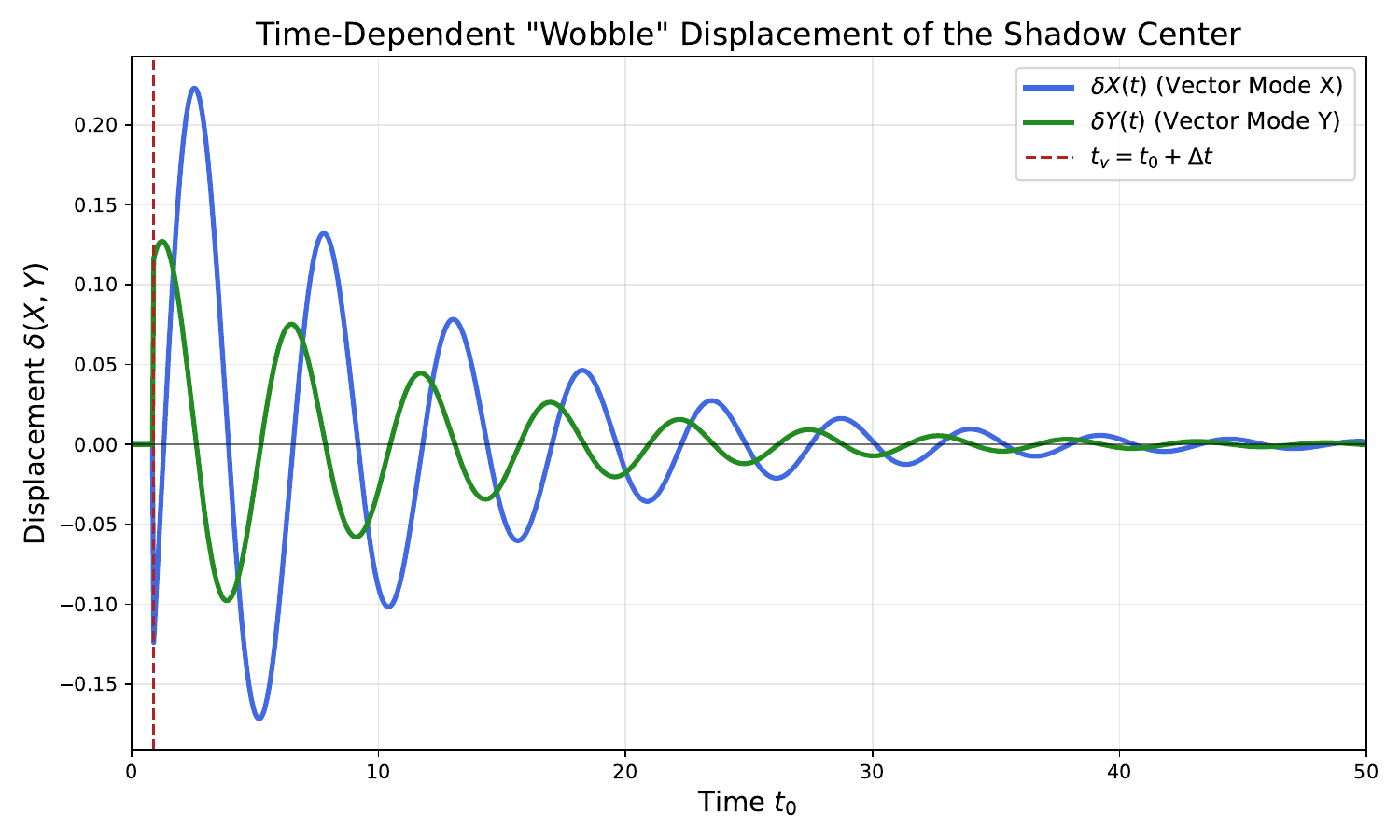}
    \caption{The translational displacement (``wobble'') of the black hole shadow center in the celestial $X$ and $Y$ coordinates. Tensor and scalar modes arrive at $t_0$ and do not shift the center, while the massive vector mode arrives at $t_v$ (delayed by $\Delta t$) and produces an asymmetric centroid shift. In General Relativity this effect is absent, and the shadow center remains fixed.}
    \label{fig:wobble_plot}
\end{figure}

By synthesizing the causal expressions derived in Eq. (\ref{eq:Area_change_expanded}) and Eqs. (\ref{eq:wobble_X_final})--(\ref{eq:wobble_Y_final}), a distinctive phenomenological signature emerges. The perturbed shadow undergoes an ordered, two-stage dynamical sequence:
\begin{enumerate}
    \item Phase I ($t = t_0$): The shadow boundary experiences an immediate, isotropic breathing oscillation $\delta A(t)$ governing its total cross-sectional area, while the shadow centroid remains fixed at the coordinate origin ($\delta X = \delta Y = 0$).
    \item Phase II ($t = t_0 + \Delta t$): Following a discrete dispersive delay $\Delta t$ determined by the vector mass $\mu_v$ in Eq. (\ref{eq:delay_exact}), the isotropic area variations decay, and the shadow undergoes an asymmetric, directional translation ($\delta X(t), \delta Y(t)$) on the celestial screen of the observer.
\end{enumerate}
This chronological sequence---an instantaneous isotropic breathing mode followed by a delayed translational wobble---acts as a macroscopic spacetime chronometer. This multi-field signature is not replicated by standard General Relativity (which lacks both modes), pure scalar-tensor theories (which lack the massive Proca carrier required for the delayed wobble), or massless vector-tensor frameworks (where $\Delta t \to 0$, causing the modes to blend indistinguishably). The identification of this sequential causal chain provides a theoretical characterization in time-domain modified gravity phenomenology, offering an explicit avenue to break both standard General Relativity and alternative-gravity parameter degeneracies simultaneously.

\section{Conclusion}
\label{sec:conclusion}

In this paper, we derived the dynamical interaction between gravitational waves and the shadow of a static modified gravity black hole \cite{Moffat:2014aja}. Using the Hamilton-Jacobi framework, we established that while the unperturbed static shadow radius, $\bar{R}_{\text{sh}}$, is governed by the deformation parameter $\alpha$, the perturbed geometry undergoes a sequence of two time-dependent modifications. First, the massless scalar field modulates the critical impact parameter of trapped photons, causing the area of the shadow to expand and contract. This yields an area fluctuation of
\begin{equation}
    \delta A(t) \approx \pi \bar{R}_{\text{sh}}^2 h_b(t),
    \label{eq:area_fluctuation_conclusion}
\end{equation}
where $h_b(t)$ is the dimensionless amplitude of the breathing mode perturbation. This volumetric breathing mode is forbidden by the volume-preserving tensor modes of General Relativity. Following this initial fluctuation, the massive vector field, $\phi_\mu$, propagating with a group velocity $v_g < 1$, arrives with a dispersive time delay $\Delta t$. This delayed wave sources secondary longitudinal metric perturbations that shift the effective angular momentum center of the photons, manifesting as an asymmetric spatial displacement in the celestial coordinates. To contextualize these dynamic geometric shifts observationally, we evaluated an extreme mass ratio inspiral \cite{Amaro-Seoane:2012vvq}. For a compact mass $m$ orbiting a supermassive host $M$ near the photon sphere, the peak localized strain reduces to the mass ratio,
\begin{equation}
    h_{\text{local}} \approx \frac{m}{M} \equiv q,
    \label{eq:mass_ratio_strain_conclusion}
\end{equation}
where $h_{\text{local}}$ is the local metric perturbation amplitude and $q$ is the dimensionless mass ratio. For a system with a ten solar mass black hole inspiring into Sagittarius A*, this yields fractional geometric deviations of $\mathcal{O}(10^{-5})$ for both the breathing area $\delta A$ and the translational displacement $\delta X$. Resolving this sequential signature provides a foundational dynamical signature for future space-based interferometry to test the existence of scalar fields and massive force carriers in the strong-field regime.

Detecting these time-dependent geometric deviations presents substantial observational challenges. The predicted fractional deviation of $\mathcal{O}(10^{-5})$ for a target such as Sagittarius A* translates to absolute dimensional changes that require microarcsecond-level angular resolution. The near-term capabilities of the next-generation Event Horizon Telescope provide an angular resolution limit of approximately $15\ \mu\text{as}$. This resolution is insufficient to directly resolve the spatial displacement induced by the delayed vector mode. Consequently, observing this foundational dynamical signature will necessitate the extended baselines provided by future space-based very long baseline interferometry missions.

Beyond purely instrumental limitations, the astrophysical environment introduces severe signal extraction complexities. The observable edge of the black hole shadow is illuminated by a turbulent, stochastic accretion flow. Separating the deterministic metric displacement from random plasma fluctuations and general relativistic magnetohydrodynamic turbulence requires advanced filtering techniques. Observers must utilize continuous, high-cadence closure phase measurements to extract the deterministic frequency $\omega_v$ from the stochastic noise floor, where $\omega_v$ represents the characteristic frequency of the massive vector field perturbation. The temporal visibilities must be sampled at a rate sufficient to capture the phase evolution of the breathing and longitudinal modes against the background astrophysical variability.

If these observational hurdles are overcome, the sequential nature of the signals provides a method to separate this modified gravity framework from other alternative theories of gravity. Pure scalar-tensor models, such as Horndeski theories, support the initial area fluctuation $\delta A(t)$ induced by the scalar breathing mode. However, without a massive vector field, the celestial center of the shadow remains fixed, yielding coordinate displacements of $\delta X = 0$ and $\delta Y = 0$. Alternatively, Einstein-Aether theories introduce vector modes, but the absence of a dispersive mass term alters the causal structure, meaning the vector perturbations arrive concurrently with the tensor modes rather than with a delayed temporal separation $\Delta t$. Therefore, the combination of a volumetric breathing mode and a delayed asymmetric displacement isolates the specific field content of scalar-tensor-vector gravity.

\begin{acknowledgements}
N.J.L. Lobos and E.T. Rodulfo gratefully acknowledge De La Salle University and the DLSU Theoretical Physics Group for their institutional support. Furthermore, we extend our sincere gratitude to the Department of Science and Technology – Accelerated Science and Technology Human Resource Development Program (DOST-ASTHRDP) for their generous and continuous support of our research endeavors.

\end{acknowledgements}

\bibliography{ref}

\end{document}